\documentclass[preprint,showpacs,showkeys,preprintnumbers,amsmath,amssymb]{revtex4}
\usepackage{dcolumn}
\usepackage{bm}
\usepackage{graphicx}
\usepackage{xcolor}

\begin{document}

\title{Hawking Radiation and $P-v$ Criticality of Charged Dynamical (Vaidya) Black Hole in Anti-de Sitter Space}

\author{Ran Li}
\email{liran@htu.edu.cn}
\affiliation{Department of Physics, Henan Normal University, Xinxiang 453007, China\\
Department of Chemistry, SUNY, Stony Brook, NY 11794, USA}

\author{Jin Wang}
\email{Corresponding Author: jin.wang.1@stonybrook.edu}
\affiliation{Department of Chemistry and of Physics and Astronomy, State University of New York at Stony Brook, Stony Brook, NY 11794-3400, USA}

\begin{abstract}

We study Hawking radiation and $P-v$ criticality of charged dynamical (Vaidya) black hole in Anti-de Sitter (AdS) space. By investigating the near horizon properties of scalar field, we derive Hawking temperature of dynamical charged (Vaidya) AdS black hole. Based on this result, by regarding the cosmological constant as the thermodynamic pressure, we investigate the analogy between the charged dynamical (Vaidya) AdS black holes in the ensemble with the fixed charge and van der Waals liquid-gas system in detail, including equation of state, $P-v$ diagram, critical point, heat capacities and critical exponents near the critical point. It is shown that the relationship among the critical pressure, critical volume, and critical temperature gets modified while the critical exponents are not affected by the dynamical nature of the black hole. We also find that, when the rate of change of the black hole horizon exceeds the critical value, the $P-v$ criticality of the charged dynamical (Vaidya) AdS black hole will disappear.

\end{abstract}

\maketitle

\section{Introduction}

Hawking and Page \cite{HawkingPage} discovered that there is a first order phase transition from the thermal AdS space to the large AdS black hole at a certain critical temperature nearly forty years ago. This phenomenon, well known as the Hawking-Page phase transition, is the first example that studies the phase transition in the black hole from the thermodynamic perspective by treating black hole as a state in the thermodynamic ensemble. Since then, the deep connection between the black hole physics in AdS and the thermodynamics have attracted much attention, especially due to the discovery of anti-de Sitter/conformal field theory (AdS/CFT) correspondence \cite{Maldacena,GKP,Witten}. In this context, the phase transitions in AdS black holes can be properly interpreted as the phase transitions in the strongly coupled system. For example, Hawking-Page phase transition was explained as the gravitational dual of the confinement/deconfinement transition in quantum chromodynamics (QCD) \cite{Wittenphase}, and the transition from the bald Reissner-Nordstr\"{o}m anti-Sitter (RNAdS) black hole to the hairy RNAdS black hole was interpreted as the normal/superconducting transition in condensed matter physics \cite{3HPRL,3HJHEP}.

In recent years, the studies of phase transition of RNAdS black hole from the thermodynamic perspective have also attracted much attention. Chamblin et al. studied the first order transition between the small and the large RNAdS black holes both in canonical ensemble and in grand canonical ensemble, and found the analogy of phase transition to the van der Waals liquid-gas system \cite{Chamblin1,Chamblin2}. Then the thermodynamic critical behavior of a RNAdS black hole in the vicinity of certain critical points was discussed by Wu in \cite{Wu}. By treating the cosmological constant as thermodynamic pressure and black hole mass as enthalpy, Dolan also discovered the remarkable analogy between van der Waals liquid-gas system and RNAdS black hole in extended phase space \cite{Dolan1,Dolan2}. Kubiznak and Mann completed this analogy by studying the critical behavior of RNAdS black hole in fixed charge ensemble \cite{Mann}. It is shown that the critical exponents coincide exactly with those of van der Waals liquid-gas system. Since then, the analogy between AdS black hole and van der Waals liquid-gas system has attracted much attention (see \cite{reviewPVcriticality} for a recent review). This analogy has been studied extensively, including the small-large black hole transition in modified gravity and higher derivative gravity \cite{Cai,WeiprdGB,Zou,RKM,Mo,Hu,Fernando}, Maxwell equal area rule of van der Waals type phase transition of charged AdS black hole \cite{Spallucci,LML}, quasinormal modes behavior near the phase transition points \cite{LZW}, van der Waals behavior of entanglement entropy in charged AdS black hole \cite{ZengLi,LiWei}, hairy black hole chemistry \cite{AMR}, Ruppeiner geometry \cite{Sahay,WeiLiuprd2019,XWY,GB}, holographic implication of black hole chemistry \cite{KR}, black hole microstructure \cite{WeiLiuPRL,WeiLiuMann}, and the alternative approach to the phase transition \cite{HPPJ,CMIM}.
In addition, the alternative phase space, where the square of the electric charge $Q$ (or more precisely $Q^s$) of black hole was regarded as a thermodynamic variable and the cosmological constant as a fixed quantity, was proposed and generalized to study the Van der Waals criticality of charged AdS black holes \cite{DSM,DS,YSD,SADD,DSM1,ZDSM,DS1,DMSM}.

However, these studies are mainly concentrated on the static or stationary black holes.
In general, the astrophysical black holes as well as primordial black holes are dynamical objects. For the realistic models, black holes are always surrounded by complex environment. So it is important to investigate the effect of environment (accreting matters) on the thermodynamic properties of black holes. Inspired by the works on the analogy between charged AdS black hole and van der Waals liquid-gas system, a question naturally arises on whether the analogy exists for the dynamical charged AdS black holes. In this paper, we will study the thermodynamics of charged dynamical (Vaidya) black hole in AdS space, and discuss the analogy to the van der Waals liquid-gas system. Vaidya black hole in AdS space is an asymptotically AdS and spherically symmetric solution to Einstein equation that describes the black hole accreting or radiating the null dust without pressure \cite{Vaidya,CZ,WW,Ghosh}. This type of black hole has been employed to explore the properties of time dependent quasinormal modes \cite{SWAS,LLQWA}, and also has been used to probe the holographic properties of quantum quench \cite{AA} and complexity \cite{Jiang}.

The purpose of the present work is to explore $P-v$ criticality of charged dynamical (Vaidya) AdS black hole and to reveal the analogy between charged dynamical (Vaidya) black hole and van der Waals liquid-gas system. We will also investigate the influence of accreting matters on the thermodynamics of Vaidya AdS black hole. For this purpose, we will firstly study the Hawking radiation of a scalar field in this background and compute Hawking temperature by employing the Damour-Ruffini method \cite{DR,ZD}. Due to the dynamical nature of Vaidya AdS black hole, the Hawking temperature of charged Vaidya AdS black hole is time dependent. This implies the thermodynamic quantities of Vaidya AdS black hole are all changing with time. However, if we investigate the system in a relative short time scale, the thermodynamic quantities can be regarded as unchanged and the system is in a local thermal equilibrium state. Then the thermodynamics and the associated Phase transition of charged Vaidya AdS black hole can be studied under this assumption. By using the result of Hawking temperature derived from Damour-Ruffini method, we are able to calculate variety of the thermodynamic properties that will reveal the $P-v$ criticality of charged dynamical (Vaidya) AdS black hole and its analogy with the van der Waals liquid-gas system, including the equation of state, the $P-v$ diagram, critical point, heat capacities, and critical exponents near the critical point, et al. Especially, it is shown that, when the rate of change of the black hole horizon exceeds a critical value, the "P-v" criticality of the charged dynamical (Vaidya) AdS black hole will disappear. This could be due to the dynamical effect caused by the accreting matters surrounding the black hole.

This paper is organized as follows. In Sec. II, we will give a brief review of charged dynamical (Vaidya) AdS black hole, and calculate its Hawking temperature by studying the near horizon's dynamics of scalar field. In Sec. III, we will discuss the $P-v$ criticality of this black hole and its analogy with van der Waals liquid-gas system. The conclusion is presented in the last section.

\section{Hawking temperature of charged dynamical (Vaidya) AdS Black hole}

We start with a review on the four dimensional charged dynamical (Vaidya) AdS black hole.
It is a solution to the Einstein field equations for a collapsing charged null fluid
in asymptotical anti-de Sitter space. The Einstein field equations are given by
\begin{eqnarray}
G_{\mu\nu}+\Lambda g_{\mu\nu}=8\pi T_{\mu\nu}\;,
\end{eqnarray}
where $\Lambda=-\frac{3}{L^2}$ is the cosmological constant.
The metric is given by
\begin{eqnarray}
ds^2=-f(v,r) dv^2 + 2cdvdr + r^2d{\Omega}^2,
\end{eqnarray}
where the metric factor $f(v,r)$ is given by
\begin{eqnarray}
f(v,r)=1-\frac{2M(v)}{r}+\frac{Q^2(v)}{r^2}+\frac{r^2}{L^2},
\end{eqnarray}
with $M(v)$ and $Q(v)$ being the arbitrary functions of $v$. $v$ plays the role of time coordinate.
The energy momentum tensor $T_{\mu\nu}$ for the charged dynamical (Vaidya) AdS black hole is given by \cite{CZ,Ghosh,lemos}
\begin{eqnarray}
T_{\mu\nu}=\rho k_{\mu}k_{\nu}+T^{m}_{\mu\nu}\;,\;\;k_{\mu}=-\delta_{\mu}^{v}\;,\;\;
k_{\mu}k^{\mu}=0\;,
\end{eqnarray}
where $T^{m}_{\mu\nu}$ is the energy momentum tensor for the electromagnetic field and $\rho$ is given by
\begin{eqnarray}
\rho=\frac{1}{4\pi r^3}\left(r\frac{dM}{dv}-Q\frac{dQ}{dv}\right)\;.
\end{eqnarray}

In Eq.(2), the parameter $c$ takes the values of $\pm 1$. To be specific, $c=1$ corresponds to the case of ingoing flow and $M(v)$ is a monotonically increasing function of the advanced time, while
$c=-1$ corresponds to the case of outgoing flow and $M(v)$ is a monotonically decreasing
function of the retarded time. For the realistic black hole in the universe, the absorption process is always dominated. We will consider the dynamical black holes that accrete matters in the present work. For this reason, we take $c=1$ in the following discussion. This is to say we are just going to discuss the influence of accreting matters on the $P-v$ criticality of dynamical black hole.

As discussed in the introduction, we firstly calculate the Hawking radiation from the charged dynamical (Vaidya) AdS black hole in terms of Damour-Ruffini method generalized by Zhao et al \cite{ZD}. Without loss of generality, let us consider the scalar field $\Psi$, where its dynamics is governed by the Klein-Gorden equation in charged dynamical (Vaidya) AdS black hole space time background
\begin{eqnarray}
\nabla_\nu \nabla^\nu \Psi-\mu^2 \Psi=0,
\end{eqnarray}
where $\mu$ is the mass of the scalar field.

By separating the variables as
$\Psi=\Phi(v,r) Y_{lm}(\theta,\phi)$, one can get the following equation after some algebra
\begin{eqnarray}
f \frac{\partial^2 \Phi}{\partial r^2} +2c \frac{\partial^2 \Phi}{\partial v \partial r}
+\frac{\partial f}{\partial r} \frac{\partial \Phi}{\partial r}
-\left[\frac{1}{r}\frac{\partial f}{\partial r}+\mu^2+\frac{l(l+1)}{r^2}\right]\Phi=0\;.
\end{eqnarray}
Near horizon behavior of the scalar field is essential in Damour and Ruffini's derivation \cite{DR}. In order to investigate the near horizon behavior of the scalar field, we introduce the generalized tortoise coordinates as
\begin{eqnarray}
r*&=&r+\frac{1}{2\kappa(v_0)}\ln(r-r_H(v))\;,\nonumber\\
v*&=&v-v_0\;,
\end{eqnarray}
where $\kappa$ is an adjustable parameter which will be determined in the following, $v_0$ is an arbitrary constant, and $r_H(v)$ is the location of the event horizon of the dynamical black hole. Without the loss of generality, we can set $v_0=0$.

Transforming into the generalized tortoise coordinates, the radial equation (7) can be casted into the following form
\begin{eqnarray}
 C_{rr} \frac{\partial^2 \Phi}{\partial r*^2} +C_{vr} \frac{\partial^2 \Phi}{\partial v\partial r*}+C_{r} \frac{\partial \Phi}{\partial r*}+ C \Phi=0\;,
\end{eqnarray}
where the coefficients are given by
\begin{eqnarray}
C_{rr}&=&f\left[1+\frac{1}{2\kappa(r-r_H)}\right]-\frac{2\dot{r}_H}{2\kappa(r-r_H)}\;,
\nonumber\\
C_{vr}&=&2\;,\nonumber\\
C_{r}&=&f'+\frac{2\dot{r}_H-f}{(r-r_H)\left[1+2\kappa(r-r_H)\right]}\;,\nonumber\\
C&=&-\left[\frac{f'}{r}+\mu^2+\frac{l(l+1)}{r^2}\right]\frac{2\kappa(r-r_H)}{1+2\kappa(r-r_H)}\;.
\end{eqnarray}

When taking the limit of $r\rightarrow r_H$, the coefficient $C_{rr}$ should take the finite value. This gives us the equation
\begin{eqnarray}
f=2\dot{r}_H\;,
\end{eqnarray}
which is nothing but the equation that determines the location of the event horizon
\begin{eqnarray}
1-\frac{2M}{r_H}+\frac{Q^2}{r_H^2}+\frac{r_H^2}{L^2}-2\dot{r}_H=0\;.
\end{eqnarray}

It is shown by Damour and Ruffini in \cite{DR} that the equation of scalar field in the near horizon region can be reduced to the standard form of wave equation. It is easy to see that
$C_{r}=C=0$ when taking the near horizon limit of $r\rightarrow r_H$. On the other hand, if we impose the condition that the coefficient $C_{rr}|_{r\rightarrow r_H}=1$, one can get \cite{ZD,SWAS,LLQWA}
\begin{eqnarray}
\kappa=\lim_{r\rightarrow r_H} \frac{f-2\dot{r}_H}{2(1-f)(r-r_H)}
=\frac{1-\frac{Q^2}{r_H^2}+\frac{3r_H^2}{L^2}-2\dot{r}_H}{2r_H(1-2\dot{r}_H)}\;,
\end{eqnarray}
where L'Hospital's Rule is used to derive this result. Then the equation (9) can be reduced to the standard form of the wave equation as expected
 \begin{eqnarray}
 \frac{\partial^2 \Phi}{\partial r*^2} +2 \frac{\partial^2 \Phi}{\partial v\partial r*}=0
\end{eqnarray}

Two linearly independent solutions of the above equation are given by
\begin{eqnarray}
\Phi_{in}&=&e^{-i\omega v}\;,\nonumber\\
\Phi_{out}&=&e^{2i\omega r*}e^{-i\omega v}\;.
\end{eqnarray}
The solution $\Phi_{in}$ represents the ingoing solution that is analytic on the horizon, while $\Phi_{out}$ represents the outgoing solution which has a logarithmic singularity at the horizon.

Damour and Ruffini suggested that the outgoing wave can be analytically extended through
the lower half complex $r$ plane into the inside of the horizon. In our present case,
the analytical continuation to extend the outgoing wave outside the bulk horizon to the outgoing wave inside the horizon can be realized by making the replacement of
 \begin{eqnarray}
 r-r_H\mapsto |r-r_H|e^{-i\pi}=(r_H-r)e^{-i\pi}\;.
 \end{eqnarray}
 Then, the outgoing wave inside the horizon by analytically continuation can be given by
 \begin{eqnarray}
 \Phi_{out}(r<r_H)=e^{\pi\omega/\kappa}e^{2i\omega r*}e^{-i\omega v}\;.
 \end{eqnarray}
 The relative probability of the scattered outgoing wave at the horizon is given by
 \begin{eqnarray}
 P=\left|\frac{\Phi_{out}(r>r_H)}{\Phi_{out}(r<r_H)}\right|^2=e^{-2\pi\omega/\kappa}\;.
 \end{eqnarray}

According to the heuristic derivation of Sannan \cite{Sannan}, we can obtain the spectral of Hawking radiation as
\begin{eqnarray}
N_{\omega}=\frac{1}{e^{\omega/T}-1}\;,
\end{eqnarray}
with the Hawking temperature
\begin{eqnarray}
T=\frac{\kappa}{2\pi}=\frac{1-\frac{Q^2}{r_H^2}+\frac{3r_H^2}{L^2}-2\dot{r}_H}{4\pi r_H(1-2\dot{r}_H)}\;.
\end{eqnarray}

Obviously, when $\dot{r}_H=0$, the result reduces to the temperature of the RNAdS black hole.
In general, $\dot{r}_H$ is non zero which represents the dynamical property of Vaidya spacetime. This expression for the charged dynamical (Vaidya) AdS black hole can give us the equation of state which is essential for the following discussion. Especially, we will take $\dot{r}_H$ as a parameter and consider the dynamical effect on the "P-v" criticality.

\section{P-v criticality of charged dynamical (Vaidya) AdS black hole}

\subsection{Equation of state}

Treating the cosmological constant as a dynamical variable was firstly proposed by Teitelboim and Brown \cite{Teitelboim,BT}. Recently, the idea of regarding the cosmological constant $\Lambda$ as a thermodynamic variable has attained increasing attention \cite{reviewPVcriticality}. This is the basic idea for the extended phase space proposal of AdS black hole, where the cosmological constant is related to the thermodynamic pressure. In the case of an asymptotically AdS black hole in four dimensions, one identifies the pressure with the cosmological constant \cite{KRT}
\begin{eqnarray}
P=-\frac{1}{8\pi}\Lambda=\frac{3}{8\pi} \frac{1}{L^2}\;.
\end{eqnarray}
With this identification, the expression Eq.(20) for Hawking temperature of charged dynamical (Vaidya) AdS black hole can be transformed into the expression for equation of state in the extended phase space.

Firstly, by substituting Eq.(21) into Eq.(20), we can get the expression of thermodynamic pressure $P$ as follows
\begin{eqnarray}
P=\frac{1-2\dot{r}_H}{2r_H}T-\frac{(1-2\dot{r}_H)}{8\pi r_H^2}+\frac{Q^2}{8\pi r_H^4}\;.
\end{eqnarray}
When $\dot{r}_H=0$, this equation can be reduced to the equation of state for the RNAdS black hole.
Comparing with the van der Waals equation, we take the specific volume $v$ as \cite{Mann}
\begin{eqnarray}
v=2r_H\;.
\end{eqnarray}
Now, we also take $\dot{r}_H=\delta$ as a parameter to study its dynamical effects on the $P-v$ criticality of charged dynamical (Vaidya) AdS black hole. Then we can obtain the equation of state in the following form
\begin{eqnarray}
P=(1-2\delta)\frac{T}{v}-\frac{(1-2\delta)}{2\pi v^2}+\frac{2Q^2}{\pi v^4}\;.
\end{eqnarray}
This equation is the equation of state for the dynamical RNAdS black hole. This provides a starting point of the following discussions.

\subsection{$P-v$ diagram and critical point}

\begin{figure}
  \centering
  \includegraphics[width=8cm]{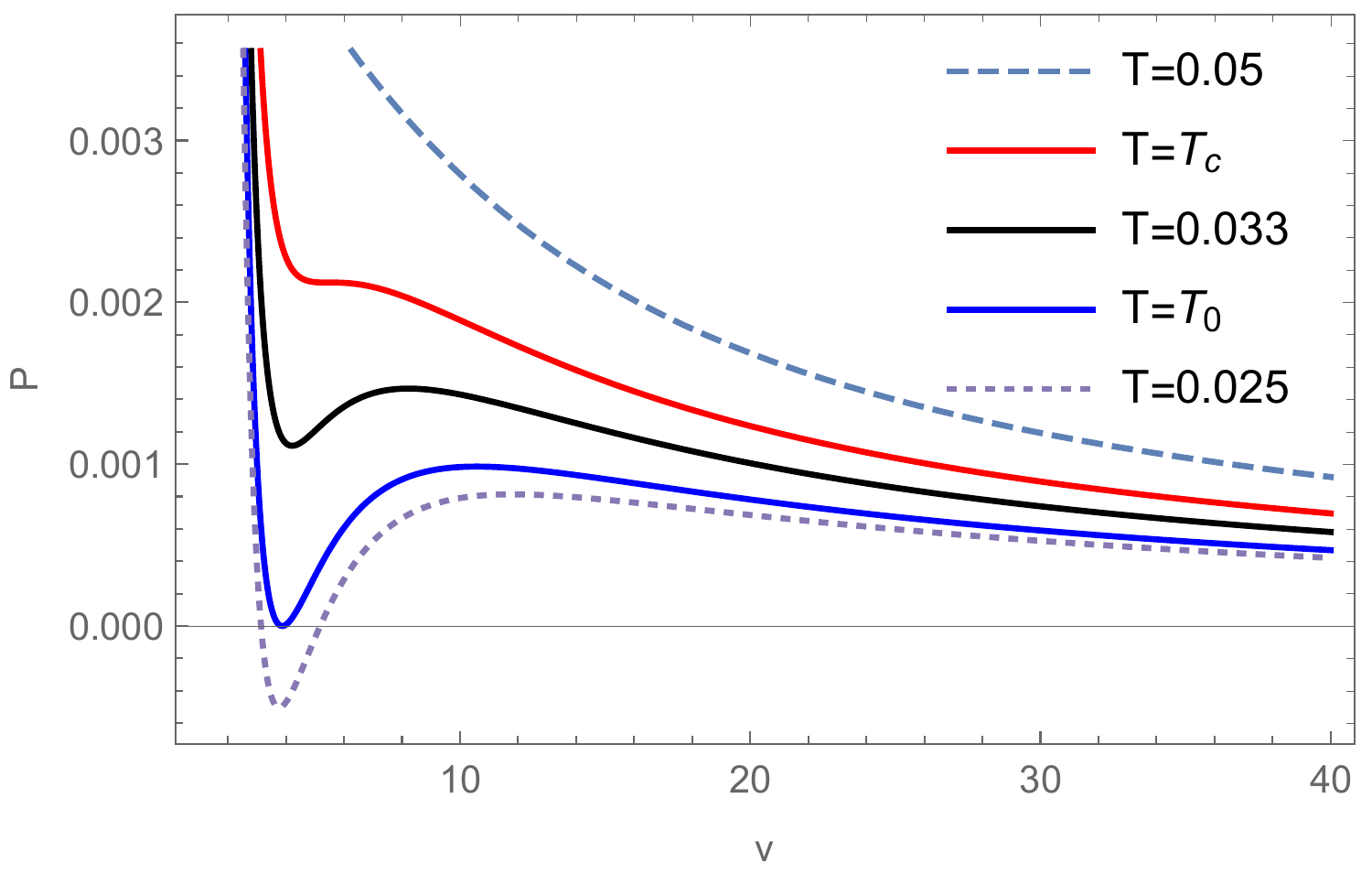}
  \includegraphics[width=8cm]{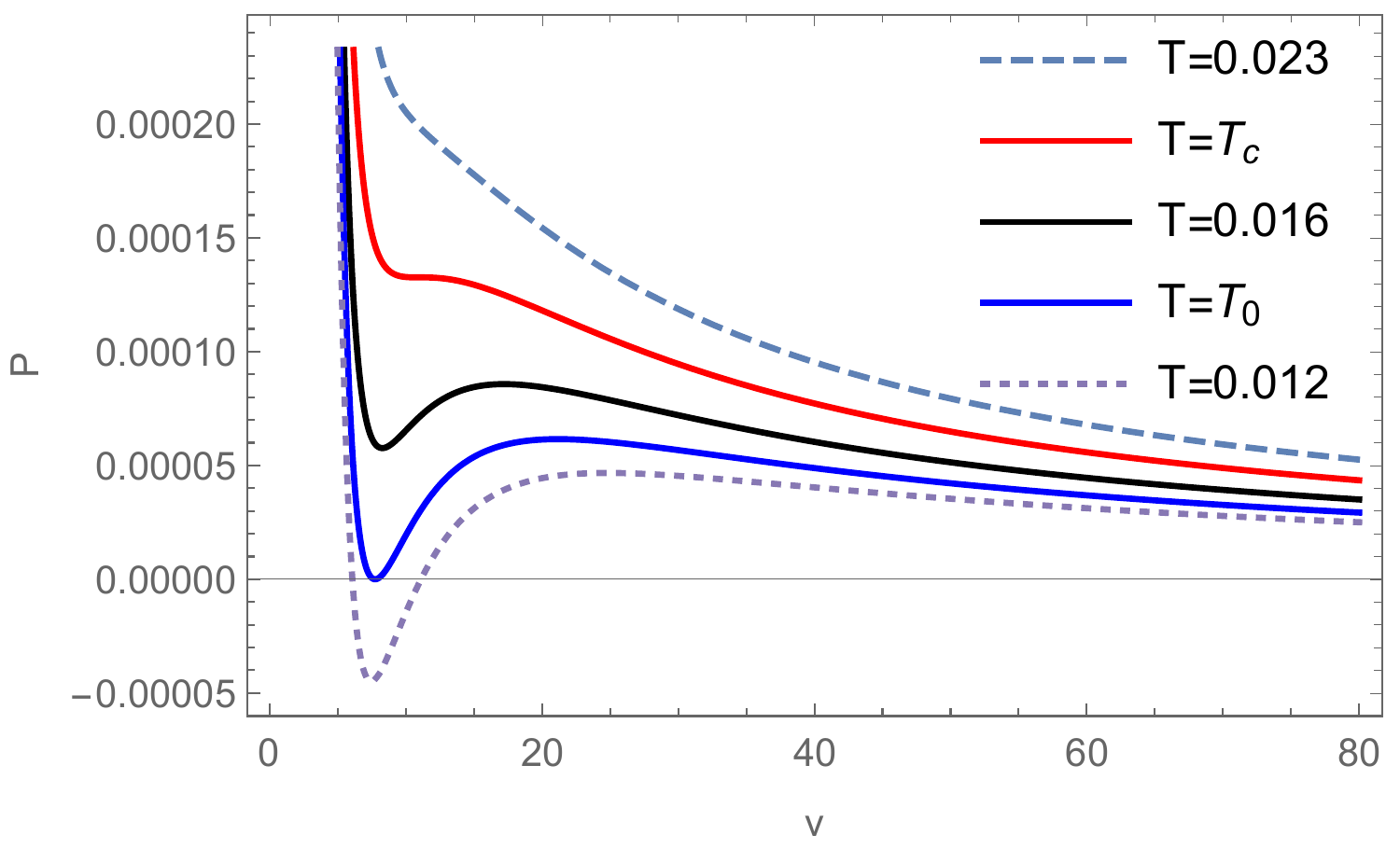}\\
  \caption{$P-v$ diagram of charged dynamical (Vaidya)AdS black hole for different parameter $\delta$. The left panel is for $\delta=0.1$ and the right one is for $\delta=0.4$. The temperature decreases from the top to the bottom. The top dashed line corresponds to the ideal gas behavior for $T>T_c$, the critical isotherm $T=T_c$ is denoted by the red solid line, the black solid line corresponds to the case of the temperature smaller than the critical temperature, and the $T=T_0$ isotherm is also displayed by the blue solid line. The dotted line is the $T<T_0$ isotherm. We have set $Q=1$ for simplicity.}
 \label{PVdiagram1}
\end{figure}

The temperature of the isotherms, i.e. the "P-v" diagrams for different values of the parameter $\delta$ are depicted in Fig.\ref{PVdiagram1}. If the parameter $\delta\geq 0.5$, the "P-V" diagram will show the unphysical behavior. Therefore, we take the parameter $\delta$ to be within the range of $0<\delta<0.5$.
Evidently, for $T=T_c$ there is an inflection point and the behaviour is similar to the Van der
Waals liquid-gas system. We mainly concentrate on the "P-v" criticality of the charged dynamical (Vaidya) AdS black hole in this subsection.

The critical point is just the inflection point in the "P-v" diagram, which is
determined by the equation
\begin{eqnarray}
\frac{\partial P}{\partial v}=\frac{\partial^2 P}{\partial v^2}=0\;,
\end{eqnarray}
which leads to
\begin{eqnarray}
T_c&=&\frac{\sqrt{1-2\delta}}{3\sqrt{6}\pi Q}\;,\nonumber\\
v_c&=&\frac{2\sqrt{6}Q}{\sqrt{1-2\delta}}\;,\nonumber\\
P_c&=&\frac{(1-2\delta)^2}{96\pi Q^2}\;.
\end{eqnarray}

Inspecting the critical values, we find the relation
\begin{eqnarray}
\frac{P_c v_c}{T_c}=\frac{3}{8}(1-2\delta)\;,
\end{eqnarray}
which is similar to the relation for the Van der Waals liquid-gas system except the factor $(1-2\delta)$. This relation of critical pressure, specific volume, and temperature is determined by the parameter $\delta$, and the relation is universally predicted for any charged dynamical (Vaidya) AdS black hole with arbitrary charge.
However, if the definition of specific volume is changed to $v=\frac{2r_H}{1-2\delta}$, this factor can be absorbed in the definition of critical specific volume. Then following the same procedure, one can easily show the relation between critical values of $P$, $v$, and $T$ recovers to the universal number $\frac{3}{8}$, which is exactly the same value for the van der Waals liquid-gas system.

From Fig.\ref{PVdiagram1}, we can see that there are three branches of black hole solutions
(i.e. small, intermediate, and large) when the temperature is smaller than the critical temperature.
The intermediate solution is unstable, thus one has to replace the oscillatory part of the isotherm by
an isobar, according to Maxwell area law. Similar to the van der Waals liquid-gas system, there is a first order phase transition from the small black hole to the large black hole.

Similar to the Van der Waals equation of state, there exists a temperature $T_0$, below which the pressure $P$ becomes negative for some $v$. At the temperature $T_0$, we have
\begin{eqnarray}
\frac{\partial P}{\partial v}=0\;,\;\;\;P=0\;.
\end{eqnarray}
The above equation can be solved to reach
\begin{eqnarray}
T_0=\frac{\sqrt{1-2\delta}}{6\sqrt{3}\pi Q}\;.
\end{eqnarray}

We can define the new parameters as
\begin{eqnarray}
p=\frac{P}{P_c}\;,\;\nu=\frac{v}{v_c}\;,\;\tau=\frac{T}{T_c}\;,
\end{eqnarray}
The equation of state can be translated into
\begin{eqnarray}
8\tau=3\nu\left(p+\frac{2}{\nu^2}\right)-\frac{1}{\nu^3}\;,
\end{eqnarray}
which is exactly the same as for the RNAdS black hole.

We have shown that the charged dynamical (Vaidya) AdS black holes have similar qualitative behaviors as the static RNAdS black holes except for the actual quantitative features. In the next subsection, we will consider the influence of the dynamical parameter $\delta$ on the analogy of van der Waals liquid-gas system to the charged dynamical (Vaidya) AdS black holes.

\subsection{Effects of accreting parameter on $P-v$ criticality}

\begin{figure}
  \centering
  \includegraphics[width=10cm]{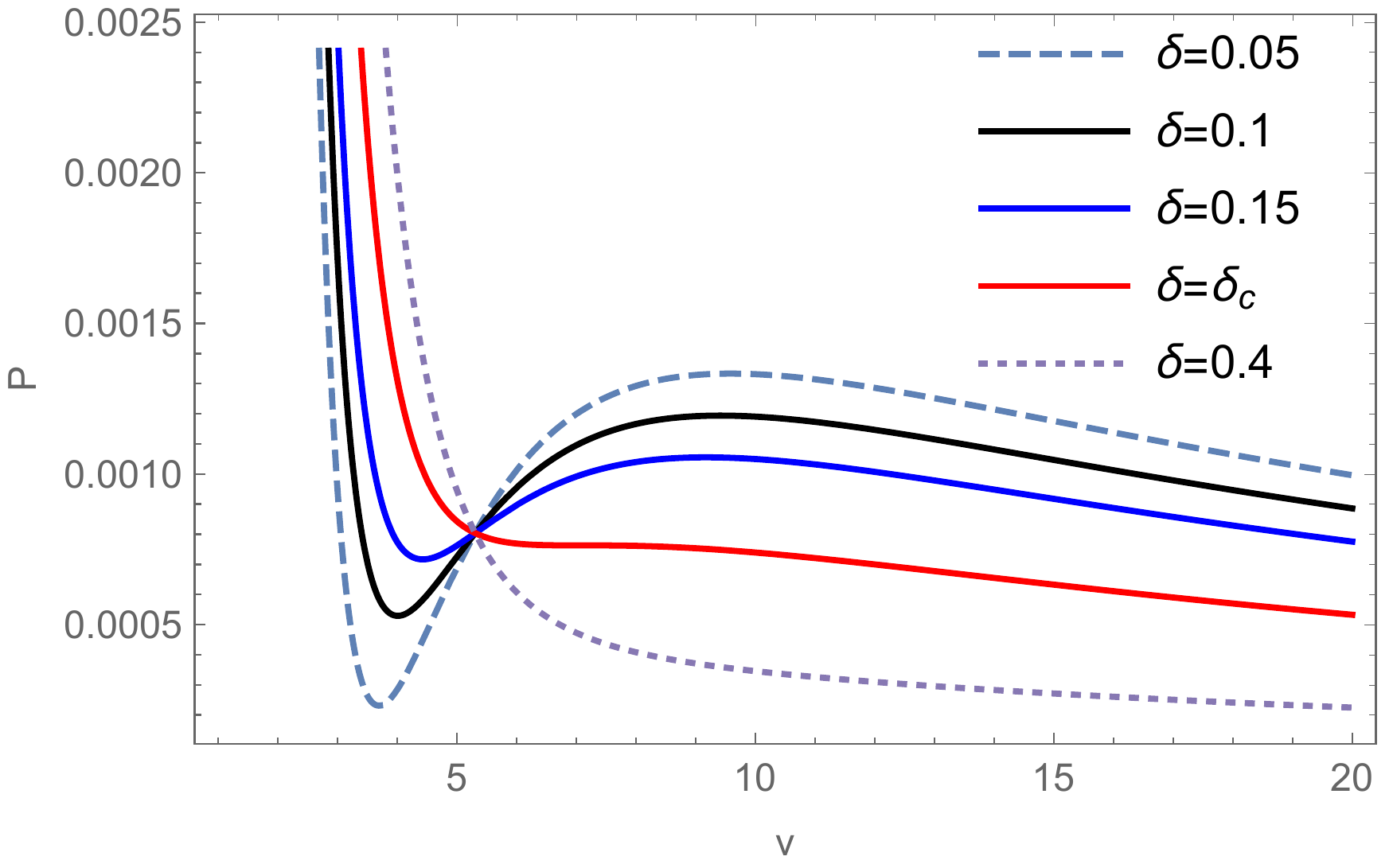}\\
  \caption{P-v diagram of the charged dynamical (Vaidya) AdS black hole for the fixed temperature $T=0.03$. When $\delta$ increases, the Van der Waals-like behavior disappears. }
  \label{antipv}
\end{figure}

If we consider the effect of the parameter $\delta$, another type of "P-v" criticality can be observed. For the fixed temperature, the "P-v" diagram of the charged dynamical (Vaidya) AdS black hole for different values of the parameter $\delta$ are depicted in Fig.\ref{antipv}.

It is observed that when the accreting parameter $\delta$ increase the van der Waals-like behavior in "P-v" diagram disappears. There is a critical value of the parameter at fixed temperature beyond which the van der Waals-like behavior disappears
\begin{eqnarray}
\delta_c=\frac{1}{2}-27\pi^2 T^2 Q^2\;,
\end{eqnarray}
which is consistent with the expression for the critical temperature in Eq.(26). Comparing the equation of state Eq.(24) with van der Waals equation $p=\frac{T}{v-b}-\frac{a}{v^2}$, we can get $b=0$, and $a=1-2\delta$. It is known that $b$ represents the volume of van der Waals molecule and $a$ is the measure of the average attraction between molecules. When increasing the parameter $\delta$, it is equivalent to decreasing the interacting force between molecules. If the parameter $\delta$ exceeds the critical value $\delta_c$, the interacting force between molecules becomes negligible, and the system behaves as an ideal gas as shown in Fig.\ref{antipv}.

Another observation from Fig.\ref{antipv} is that there exists a fixed point. The emergence of the fixed point in these curves may be just a consequence of the mathematical structure of the equation of state. The fixed point which is not changed when varying the parameter $\delta$ can be determined by the following equation
\begin{eqnarray}
\frac{\partial P}{\partial \delta}=0\;,
\end{eqnarray}
which gives the analytical results as
\begin{eqnarray}
v_f&=&\frac{1}{2\pi T}\;,\nonumber\\
p_f&=&32\pi^3 T^4 Q^2\;.
\end{eqnarray}

These behaviors of the equation of state are dramatically different from the behavior of the static RNAdS black holes. In fact, the parameter $\delta$ represents the dynamical effect of vaidya spacetime. It can be naturally concluded that these behaviors are triggered by the dynamical effect of accreting matters around black hole.

\subsection{Heat capacities}

In this subsection, we will calculate the heat capacities of the charged dynamical (Vaidya) AdS black hole and discuss their behaviors. It is well known that the entropy is given in terms of the area of the event horizon
\begin{eqnarray}
S=\pi r_H^2\;.
\end{eqnarray}
The thermodynamic volume can be computed as
\begin{eqnarray}
V=\left(\frac{\partial M}{\partial P}\right)_{S,Q}=\frac{4}{3}\pi r_H^3\;,
\end{eqnarray}
where the black hole mass can be obtained from Eq.(12) by using the relation between the thermodynamic pressure and the cosmological constant. From the expression of entropy, one can obtain
\begin{eqnarray}
C_V=T\left.\frac{\partial S}{\partial T}\right|_{V}=0\;.
\end{eqnarray}

By using the relation
\begin{eqnarray}
T=\frac{1}{4(1-2\delta)\sqrt{\pi S}}\left((1-2\delta)-\frac{\pi Q^2}{S}+8PS\right)\;,
\end{eqnarray}
which is obtained from Eq.(20) by replacing horizon radius with entropy,
the heat capacity $C_P$ is given by
\begin{eqnarray}
C_P=T\left.\frac{\partial S}{\partial T}\right|_{P}
=2S\frac{8PS^2+(1-2\delta)S-\pi Q^2}{8PS^2-(1-2\delta)S+3\pi Q^2}\;.
\end{eqnarray}

\begin{figure}
  \centering
  \includegraphics[width=8cm]{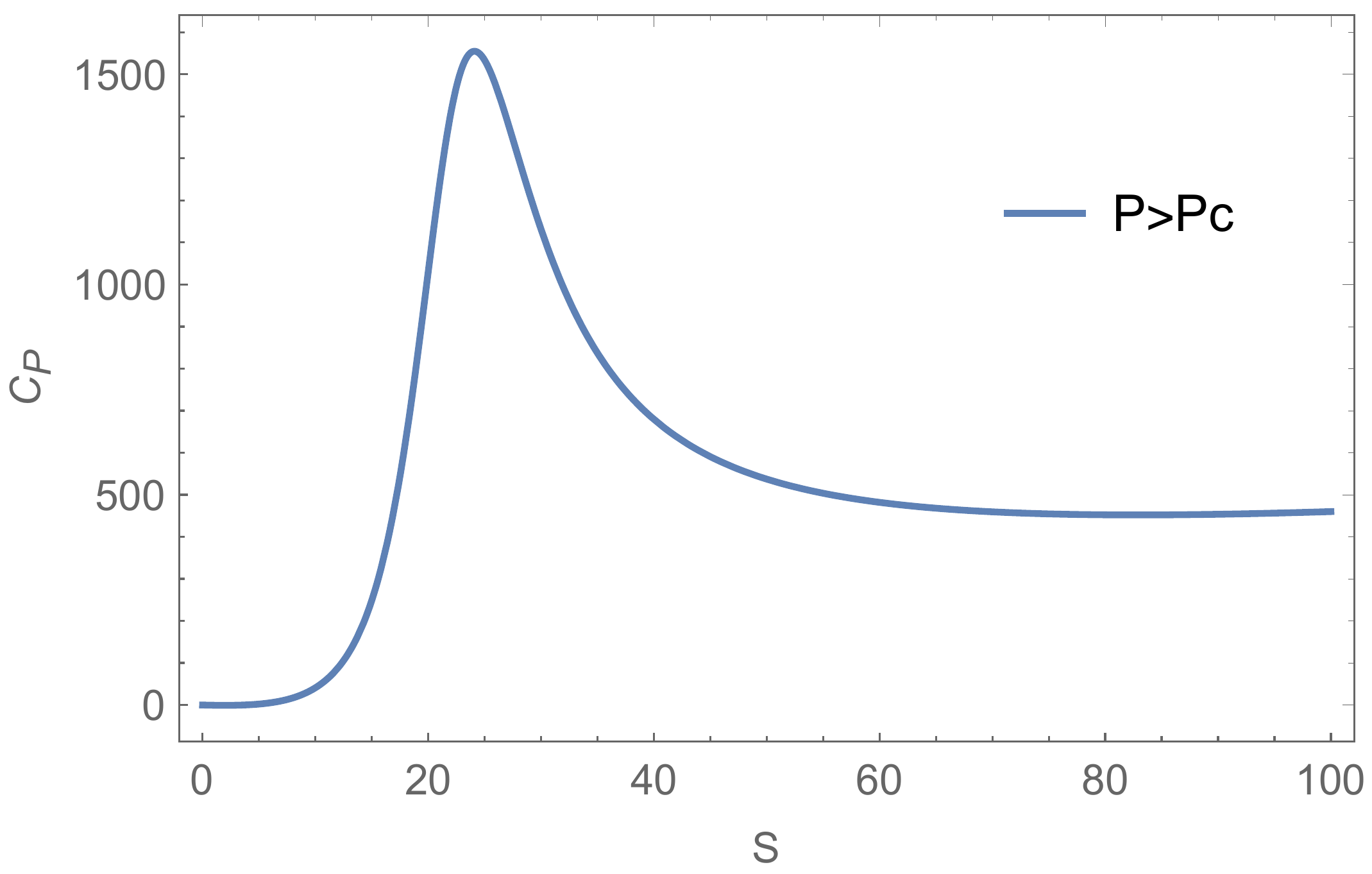}\\
  \includegraphics[width=8cm]{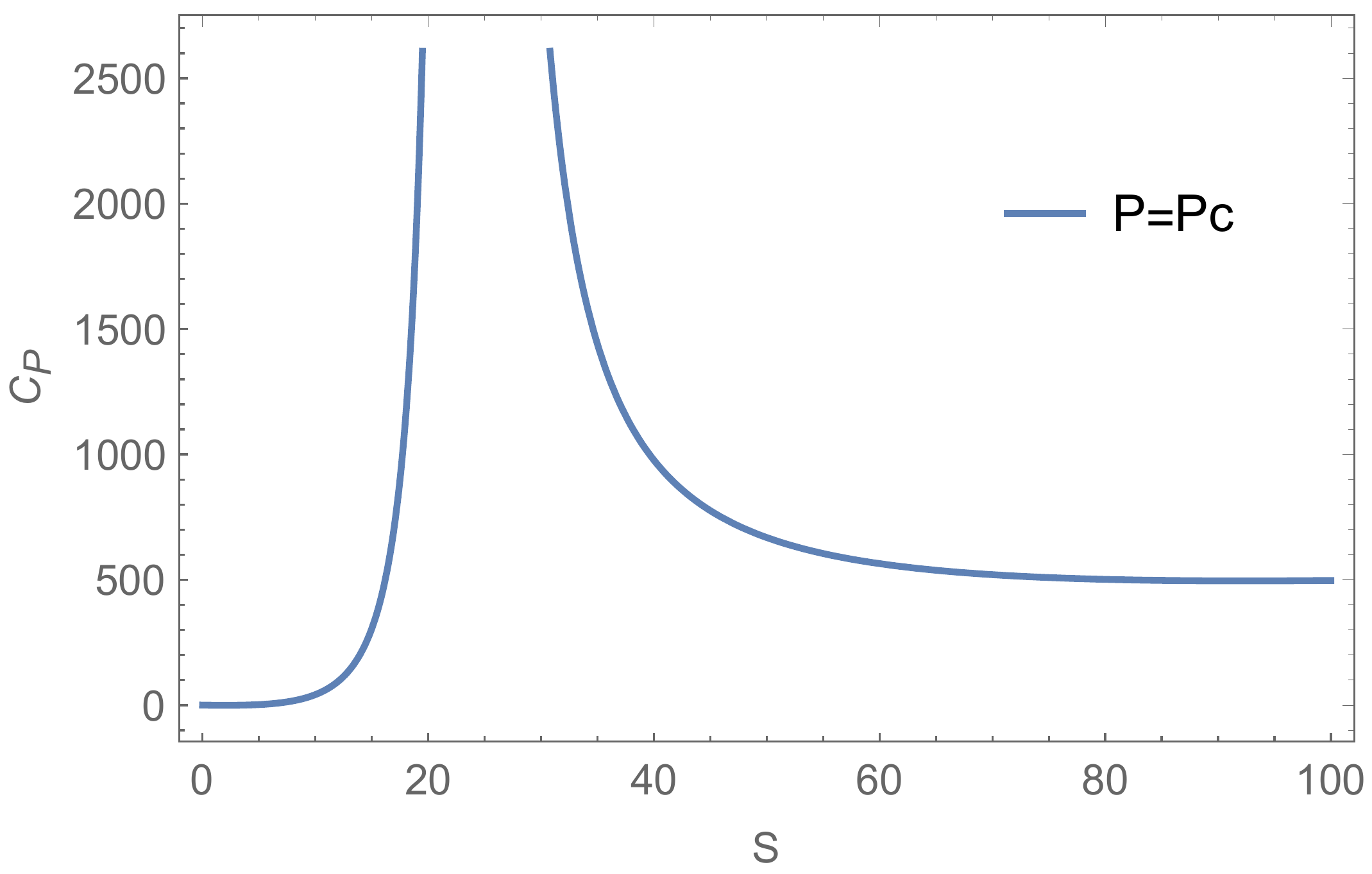}\\
   \includegraphics[width=8cm]{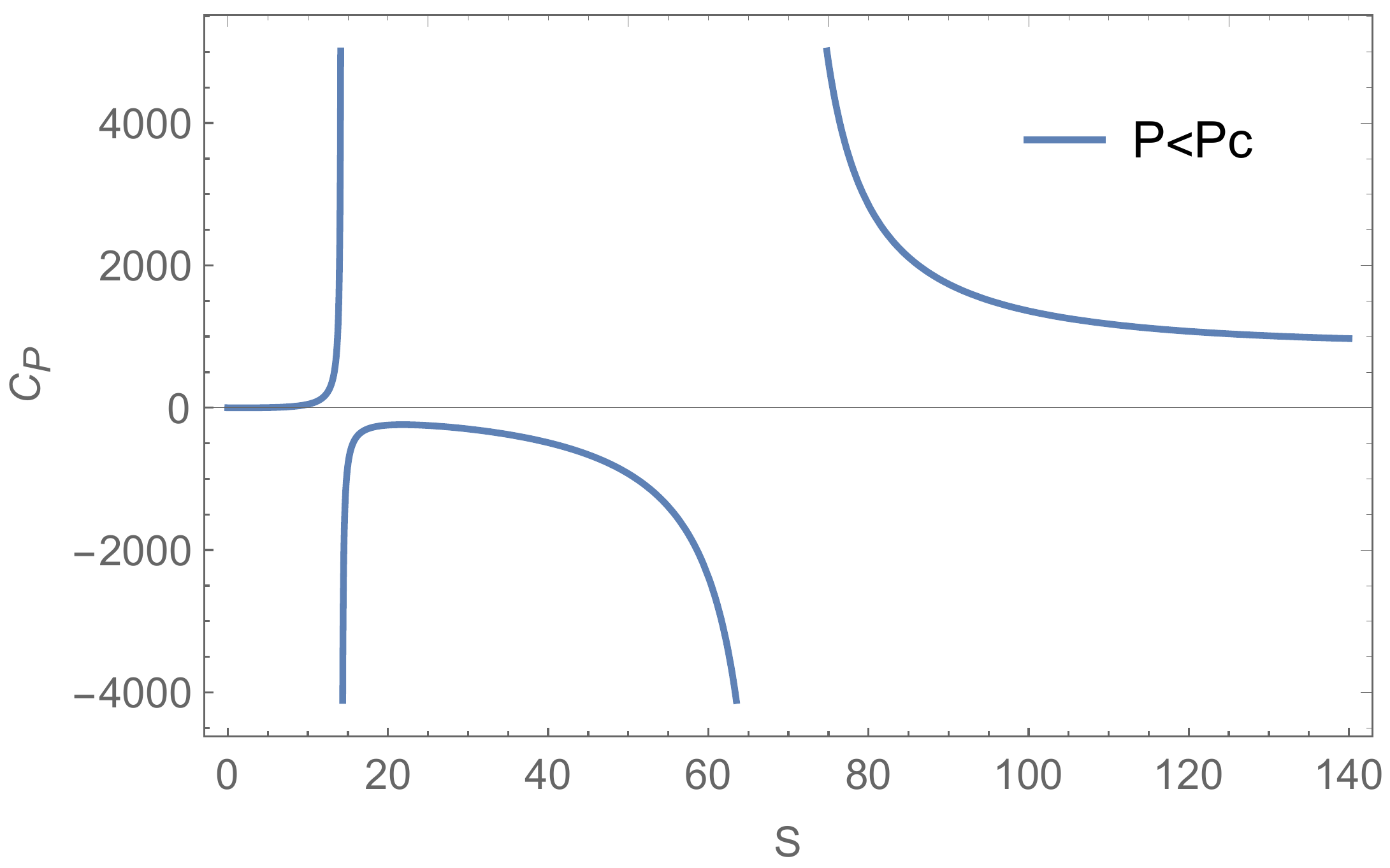}\\
  \caption{Heat capacity $C_P$ as a function of entropy $S$ when $P>P_c$, $P=P_c$, and $P<P_c$.}
  \label{Cpdiagram}
\end{figure}

The local stability of the black holes can be discussed from the behavior of the heat capacity.
We have plotted the heat capacity as a function of entropy $S$ in Fig.\ref{Cpdiagram} for different pressures. When $P>P_c$, it can be seen from the top panel in Fig.\ref{Cpdiagram} that the heat capacity has no singular point and is always positive except for small values of $S$. The region where $C_P<0$ indicates the instability of the black hole. This implies that the small black holes are always locally unstable.

In the middle panel of Fig.\ref{Cpdiagram}, we show the behavior of the heat capacity at the critical point. It can be seen that the heat capacity is divergent at the critical point. The divergence of the heat capacity is a signature of the first-order phase transition. Therefore, we can conclude the type of phase transition of the small/large black holes is first-order. 
 
From the analytical expression for the heat capacity in Eq.(38), the specific heat $C_P$ becomes singular at $8PS^2-(1-2\delta)S+3\pi Q^2=0$, and since
$S=\pi r_H^2$ this occurs when
\begin{eqnarray}
8\pi P r_H^4-(1-2\delta) r_H^2+3 Q^2=0\;,
\end{eqnarray}
which is exactly at the critical point determined by the equation of state.

When the pressure is smaller than the critical value, there are two singular points for the heat capacity as shown in the bottom panel in Fig.\ref{Cpdiagram}. The two singular points divide the curve of heat capacity into three branches, which correspond to the small black hole, intermediate black hole, and large black hole, respectively. This plot indicates that the small black hole and the large black hole are stable, while the intermediate black hole is unstable. This conclusion also implies that the oscillatory part of the isotherm in "P-v" diagram should be replaced by an isobar according to Maxwell's area law. It should also be noted that, in the middle panel and bottom panel in Fig.\ref{Cpdiagram}, there are small regions where $C_P<0$ for the small black hole.

\subsection{Critical exponents}

In this subsection, we will calculate the critical exponents which describe the behavior of the physical quantities near the critical point.
Firstly, we introduce the new parameters as follows
\begin{eqnarray}
t=\frac{T}{T_c}-1=\tau -1\;,\;\;
\omega=\frac{V}{V_c}-1\;.
\end{eqnarray}

There are four critical components, $\alpha$, $\beta$, $\xi$, and $\tilde{\delta}$, which are defined as follows:

Critical exponent $\alpha$ governs the behavior of the specific
heat at constant volume as:
\begin{eqnarray}
C_V\propto |t|^{-\alpha}\;.\nonumber
\end{eqnarray}

Critical exponent $\beta$ describes the behavior of the order
parameter $\eta=V_l-V_s$ on the given isotherm
\begin{eqnarray}
\eta\propto |t|^{\beta}\;.\nonumber
\end{eqnarray}

Critical exponent $\xi$ governs the behavior of isothermal compressibility $\kappa_T$ defined
as:
\begin{eqnarray}
\kappa_T\propto |t|^{\xi}\;.\nonumber
\end{eqnarray}

Critical exponent $\tilde{\delta}$ governs the behavior of pressure relative to the volume as:
\begin{eqnarray}
P-P_c\propto (V-V_c)^{\tilde{\delta}}\;.\nonumber
\end{eqnarray}

From the discussion in the above subsection, we know that $C_V=0$.
Hence, the critical exponent $\alpha=0$.

We can rewrite the equation of state as
\begin{eqnarray}
p=\left(\frac{8(1+t)}{3(1+\omega)^{1/3}}-\frac{2}{(1+\omega)^{2/3}}
+\frac{1}{3(1+\omega)^4/3}\right)\;.
\end{eqnarray}
We approximate the equation of state (42) near the critical phase transition as
\begin{eqnarray}
p=\left(1+\frac{8}{3}t-\frac{8}{9}t\omega-\frac{4}{81}\omega^3
+O(t\omega^2,\omega^4)\right)\;.
\end{eqnarray}
Differentiating this series for a fixed $t < 0$ we obtain
\begin{eqnarray}
dp=\left(-\frac{8}{9}t-\frac{4}{27}\omega^2\right)d\omega\;.
\end{eqnarray}
During the phase transition from the small black hole to the large black hole,
the pressure is constant, i.e. $p_l=p_s$ (constant cosmological constant), one can obtain
\begin{eqnarray}
\left(1+\frac{8}{3}t-\frac{8}{9}t\omega_l-\frac{4}{81}\omega_l^3\right)=
\left(1+\frac{8}{3}t-\frac{8}{9}t\omega_s-\frac{4}{81}\omega_s^3\right)\;.
\end{eqnarray}
Then, by employing Maxwell equal area law, we obtain the following equation
\begin{eqnarray}
\int_{\omega_l}^{\omega_s} \omega dp =\int_{\omega_l}^{\omega_s}  \omega\frac{dp}{d\omega} d\omega=0\;.
\end{eqnarray}
By substituting Eq.(44) into Eq.(46) and performing integrations, we obtain the following relation
\begin{eqnarray}
12 t \omega_l^2+\omega_l^4=12 t \omega_s^2+\omega_s^4\;,
\end{eqnarray}
which results in a unique non-trivial solution
\begin{eqnarray}
\omega_l=-\omega_s=3\sqrt{-2t}\;.
\end{eqnarray}
The behavior of the order parameter $\eta$ can be expanded as
\begin{eqnarray}
\eta=V_c (\omega_l-\omega_s)=2V_c \omega_l=6V_c \sqrt{-2t}\propto(-t)^{1/2}\;.
\end{eqnarray}
So we conclude that $\beta=\frac{1}{2}$.

By differentiating Eq.(53) we obtain
\begin{eqnarray}
\left.\frac{\partial P}{\partial V}\right|_{T}=\frac{P_c}{V_c}\frac{\partial p}{\partial \omega}
=-\frac{8}{9}\frac{P_c}{V_c}t+O(\omega)\;.
\end{eqnarray}
Hence, the behavior of isothermal compressibility near the critical point is given by
\begin{eqnarray}
\kappa_T=-\frac{1}{V}\left.\frac{\partial V}{\partial P}\right|_T
\propto t^{-1}\;.
\end{eqnarray}
So the critical exponent $\xi=1$.

Finally, the shape of the critical isotherm at $t=0$ is given by
\begin{eqnarray}
p\approx 1-\frac{4}{81}\omega^3\;.
\end{eqnarray}
Therefore the critical exponent $\tilde{\delta}=3$\;.

From the above calculations, we have obtained all the critical exponents in the P-v criticality of
charged dynamical (Vaidya) AdS black hole in the extended phase space by treating the cosmological constant as a pressure. It is clear they are essentially the same as those for the RNAdS black hole.
Therefore, we can conclude that for the low accretion d the critical behavior of the charged dynamical (Vaidya) AdS black hole is in the same universality class as the charged AdS black hole. It is easy to check that these critical exponents satisfy the following thermodynamic scaling laws
\begin{eqnarray}
\alpha+2\beta+\xi=2,\nonumber\\
\alpha+\beta(1+\tilde{\delta})=2\;,\nonumber\\
\xi(1+\tilde{\delta})=(2-\alpha)(\tilde{\delta}-1)\;,\nonumber\\
\xi=\beta(\tilde{\delta}-1)\;.
\end{eqnarray}
The critical exponents associated with the P-v criticality
of the charged dynamical (Vaidya) AdS black hole are basically the same as those
in the van der Waals liquid-gas system, and they obey similar scaling laws.

\section{Conclusion}

In this paper, we firstly studied the Hawking radiation of charged dynamical (Vaidya) AdS black hole and obtained the corresponding Hawking temperature. It is shown that, by properly approaching the near horizon region of black hole, the equation of motion for the scalar field can be reduced to the standard wave equation in two dimensions. The ingoing and the outgoing solutions to the wave equation  give rise to the radiation probability and the radiation spectrum of the black hole. Because of the dynamical nature of charged dynamical (Vaidya) AdS black hole, Hawking temperature is time dependent.

Then, we investigated the "P-v" criticality of charged dynamical (Vaidya) AdS black hole in extended phase space by identifying the cosmological constant with the thermodynamic pressure. In this framework, black hole is considered as a state in the canonical ensemble. By fixing the charge $Q$ and taking the rate of the change of the black hole horizon as a parameter, we have investigated various aspects that reveal the analogy between charged dynamical (Vaidya) AdS black hole and van der Waals liquid-gas system in details, including equation of state, "P-v" diagram, critical point, heat capacities, and the critical exponents near the critical point. We also consider the effect of accreting matters to the "P-v" criticality of black hole. We find that, when the parameter $\delta$ representing the rate of change of black hole horizon exceeds the critical value, the "P-v" criticality of charged dynamical (Vaidya) AdS black hole will disappear.

\section*{Acknowledgement}

RL is partially supported by the key research project of
universities in Henan province under Grant No. 18A140005.

\end{document}